\documentclass[12pt]{iopart}

\usepackage{color}
\usepackage[dvips]{graphicx}

\def\bfk{\mathbf{k}}
\def\bfkF{\mathbf{k}_F}
\newcommand{\NdLSCO}{La$_{1.48}$Nd$_{0.4}$Sr$_{0.12}$CuO$_4$}
\newcommand{\LBCOx}{La$_{2-x}$Ba$_{x}$CuO$_4$}

\begin{document}

   \title{
Electronic structure near the 1/8-anomaly in La-based cuprates
      } 
      
    \author{J.\ Chang$^1$, Y. Sassa$^1$, S. Guerrero$^2$, M.\ M\aa nsson$^{1,3}$, M.\ Shi$^4$, S.\ Pailh\'es$^1$, A. Bendounan$^1$, R.\ Mottl$^1$, T.\ Claesson$^3$, O.\ Tjernberg$^3$, L.\ Patthey$^4$, M.\ Ido$^5$, M.\ Oda$^5$, N.\ Momono$^{5,6}$, C. Mudry$^2$ and J.\ Mesot$^{1,7}$}
\address{$^1$ Laboratory for Neutron Scattering, 
             ETH Zurich and PSI Villigen, CH-5232 Villigen PSI, Switzerland}
\address{$^2$ Condensed Matter Theory Group, 
             Paul Scherrer Institute, CH-5232 Villigen PSI, Switzerland}
\address{$^3$ Materials Physics, Royal Institute of Technology KTH, S-164 40 Kista, Sweden}		     
\address{$^4$ Swiss Light Source, Paul Scherrer Institute, CH-5232 Villigen PSI, Switzerland}
\address{$^5$ Department of Physics, Hokkaido University - Sapporo 060-0810, Japan}
 \address{$^6$ Department of Materials Science and Engineering, Muroran Institute of
 Technology, Muroran 050-8585, Japan} 
 \address{$^7$ Institut de la materi\`ere complexe, Ecole Polytechnique Fed\'ed\'erale de Lausanne (EPFL), CH-1015 Lausanne, Switzerland}

\date{\today}
%

\begin{abstract}
We report an angle resolved photoemission study of the electronic structure of the pseudogap state in \NdLSCO\ ($T_c<7$ K). 
Two opposite dispersing Fermi arcs are the main result of this study. 
The several scenarios that can explain this observation are discussed.  
\end{abstract}
\submitto{\NJP}
\maketitle
\section{Introduction}
Deciphering the microscopic mechanism responsible for high-temperature (high-$T_c$) 
superconductivity has remained an elusive challenge for 20 years. 
Progress has been held back by the difficulty to characterize and understand
the all but conventional normal states from which high-$T_c$ superconductivity 
emerges in the underdoped and optimally doped regimes, respectively.
For example, the pseudogap state --
the normal state of underdoped high-$T_c$ superconductors -- 
is characterized by two properties that are difficult 
to reconcile.
First, angular resolved photoemission spectroscopy (ARPES) studies of the 
electronic structure in the pseudogap state have revealed 
the existence of hole-like arcs centered 
around the  diagonal of the Brillouin zone
supporting gapless quasiparticles%
~\cite{normannature98,shenscience05,kanigelnature06,hossain08}.
Fermi arcs 
cannot be attributed to a Fermi liquid (FL): 
the Fermi surface of a FL 
can only terminate at the boundary 
of the Brillouin zone.
Second, transport measurements in high-magnetic field 
have revealed the existence of quantum oscillations
in the pseudogap state of YBa$_2$CuO$_{7+\delta}$~\cite{leyraudnature07,yellandarxiv07,banguraarxiv07,jaudetarxiv07} (YBCO).
Quantum oscillations come about
when quasiparticles are orbiting around a closed Fermi surface.
These and other hallmarks of the pseudogap state
has lead to two conflicting interpretations; either the pseudogap 
state is a precursor to superconductivity or it is related 
to an order that competes with superconductivity~\cite{norman05,lee08,kivelsonRMP03}.

Here we show by ARPES 
that the electronic structure 
in the pseudogap state of \NdLSCO\ ($T_c<7K$)
consists of two oppositely dispersing arcs. 
We present furthermore a systematic study of the 
normal state electronic structure of Nd-free La$_{2-x}$Sr$_x$CuO$_4$
near the 1/8-anomaly.
The possible origins of the second arc is discussed and we show 
that it can be naively modeled by assuming the existence of a
Fermi pocket.
  
\section{Methods}
Our \NdLSCO\ sample, grown by the traveling-solvent floating-zone method,
has $T_c<7$ K. Previous $\mu$SR and neutron diffraction experiments
on the same sample are published in Refs.%
~\cite{christensenprl07} and \cite{changarxiv07}.
The ARPES experiments were performed at the SIS-beamline of the Swiss Light Source (SLS)
at the Paul Scherrer Institute (PSI) using 
55 eV circular polarized light. The overall momentum resolution
was 0.15 degrees and the pseudogap was measured with an energy resolution 
of $\sim 18$ meV. The measurements were preformed under ultra-high vacuum 
condition and the samples were cleaved at the temperature $T=15$ K.
(The lowest accessible temperature on this instrument, $T\approx 10$ K, 
remains above $T_c$.)
The ARPES data presented here were recorded in the second Brillouin 
zone but presented in the first zone for convenience. The Fermi level
was determined from a  spectrum  recorded on polycrystalline copper
on the sample holder.
      
\section{Results}

\subsection{Nodal and anti-nodal spectrum}

\begin{figure*}
\begin{center}
\includegraphics[width=0.99\textwidth]{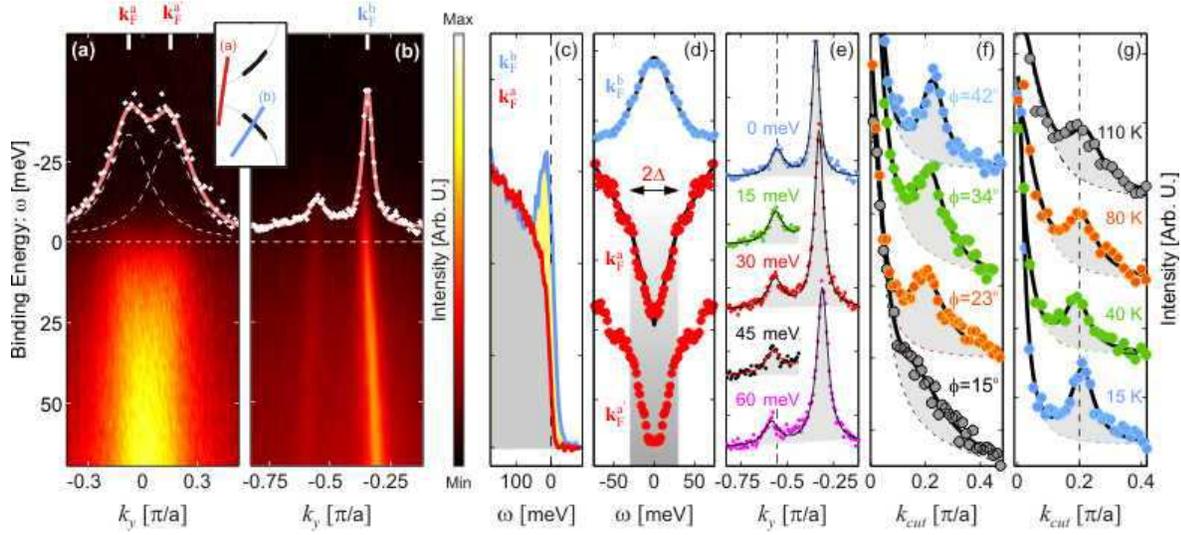}
\caption{
(a)-(b) ARPES intensity measured at $T=15$ K as a function of binding energy 
$\omega$ and momentum $\mathbf{k}$ along the red and blue cuts shown 
in the top inset of (b). The ratio of maximum intensity 
between  (a) and (b) is $\sim3/4$. White points 
are the momentum distribution curves (MDC) at $\omega=E_F$ 
and the solid lines are Lorentzian fits with a linear background. 
The double-peak structure stems from the cuts that are crossing two branches. 
(c) Energy distribution curves (EDC) at 
$\bfk=\bfkF^a$ (red) and $\bfk=\bfkF^b$ (blue) for 
the spectra shown in (a) and (b). 
(d) Symmetrized EDC for $\mathbf{k}=\mathbf{k}_{F}^{a,a^{\prime},b}$. 
The solid black lines are guides to the eye. 
(e) MDC at $\omega=0,15,30,45$, and 60 meV for the spectra shown in (b).
To avoid overlapping of the primary branches, 
only the secondary branch is
displayed when $\omega=15$ and 45 meV. 
(f) MDC at $\omega=3$ meV for four different cuts, 
showing the secondary branch as a function of the Fermi surface angle $\phi$
defined in Fig.~\ref{fig:fig2}.
(g) Temperature dependence of the MDC at $\omega=E_F$ for 
cut (b).
For both (f) and (g) the peak position of each primary branch is centered at zero 
momentum.
The  solid lines are Lorentzian fits to a double-peak structure, 
while the dashed lines show the shape of a single Lorentzian. 
   }
\label{fig:fig1}
\end{center}
\end{figure*}

\begin{figure}
\begin{center}
\includegraphics[width=0.44\textwidth]{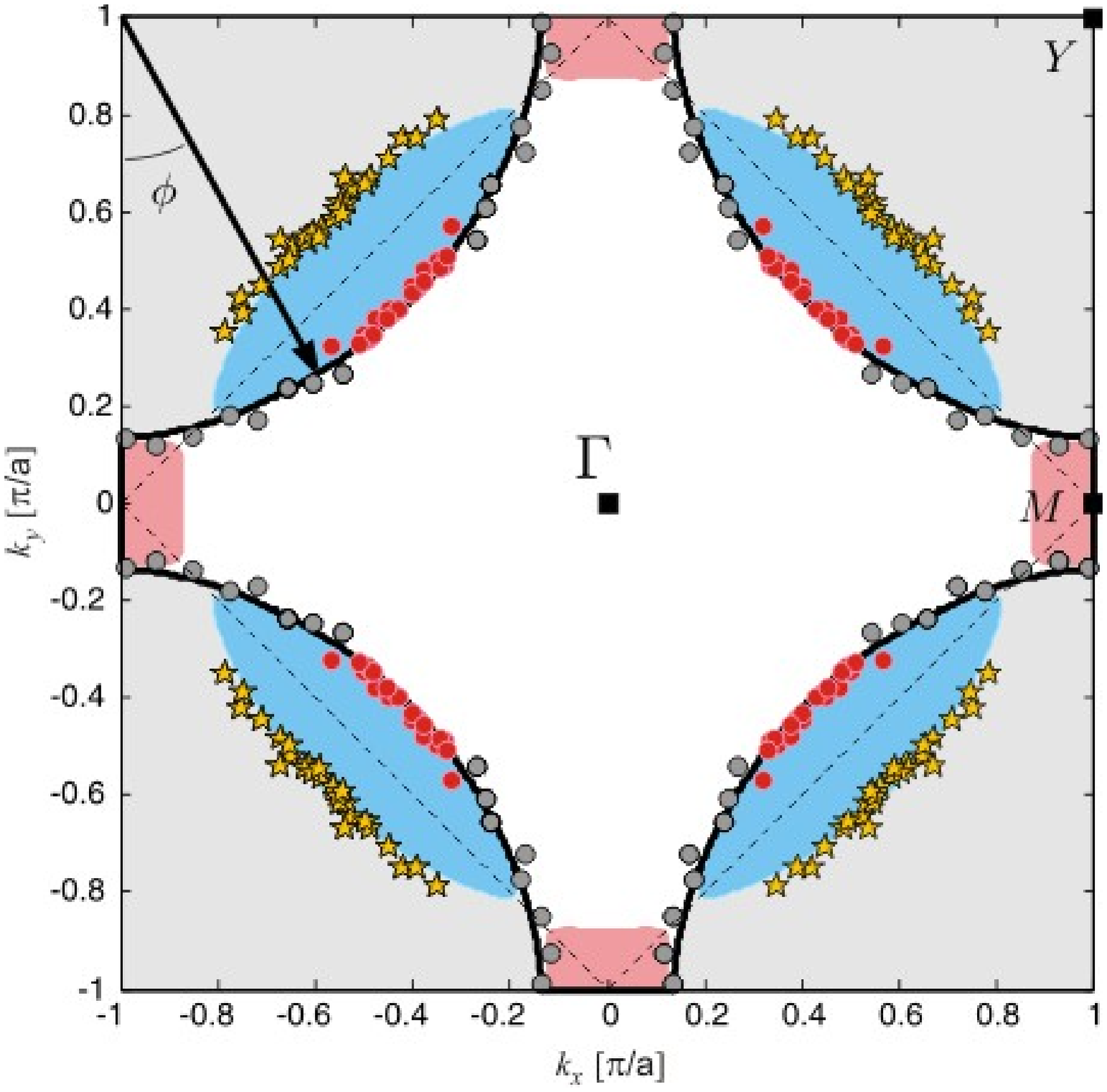}
\caption{
The locations of the peaks in the MDC at $E_F$ define the
\textit{underlying Fermi surface} and are denoted by circles
and stars. Momenta along the underlying Fermi surface characterized by
gapless quasiparticles are denoted by red circles and grey circles denote 
the part of the Fermi surface where the quasiparticles are suppressed by 
the pseudogap. The data have been symmetrized with respect to
the tetragonal crystal structure.
The solid black line stems from the Fermi surface derived from
Eq.~(\ref{eq:eq1}) while the blue and pink colored regions
depict the hole and electron pockets predicted from Eq.~(\ref{eq:eq2}).
The dashed lines bound the reduced Brillouin zone.
        }
\label{fig:fig2}
\end{center}
\end{figure}

Figures~\ref{fig:fig1}(a) and \ref{fig:fig1}(b) 
show two typical ARPES spectra collected 
close to the M-Y (zone boundary/anti-nodal) and $\Gamma$-Y 
(zone diagonal/nodal) directions (see inset)
in an incommensurate antiferromagnetic and charge ordered phase 
(the so-called stripe phase) of \NdLSCO\ (NdLSCO) at $T=15$ K, {\it i.e.}, 
well above the superconducting $T_c$.
The ARPES intensity is displayed in a false color scale 
as a function of binding energy $\omega$ and momentum $\mathbf{k}$
as indicated in the inset. 
The spectra can be analyzed with the help of constant-momentum cuts 
called energy distribution curves (EDC) or constant-energy cuts 
called momentum distribution curves (MDC).
The white points in Figs.~\ref{fig:fig1}(a) and \ref{fig:fig1}(b) 
define MDC at the Fermi level $E_F$ that can be fitted by 
(double) Lorentzian's with a linear background.
In doing so, the characteristic momenta $\mathbf{k}_F$ depicted by 
circles and stars in Fig.~\ref{fig:fig2} can be identified. 
This set of $\mathbf{k}_F$ defines
the \textit{underlying Fermi surface} shown in Fig.~\ref{fig:fig2}.

In Fig.~\ref{fig:fig1}(c) we contrast the EDC at 
$\mathbf{k}=\mathbf{k}_F^a$ and $\mathbf{k}=\mathbf{k}_F^b$,
{\it i.e.}, for parts of the underlying Fermi surface close to the 
 boundary and  diagonal 
of the Brillouin zone, respectively.
An intense spectral peak with the leading edge reaching
$E_F$ is observed along cut (b)
whereas no spectral peak is visible in the spectrum of cut (a).
Moreover, the leading edge along cut (a) is shifted 
away from $E_F$, thereby revealing the existence of a gap $\Delta$ 
consistent with the pseudogap observed in the normal state of 
underdoped cuprates. We used the so-called symmetrization method 
to factor out the Fermi distribution%
~\cite{normannature98}. 
As in strongly underdoped 
Bi$_2$Sr$_2$CaCuO$_2$ (Bi2212) \cite{tanakascience06},
our symmetrized EDC exhibit inflection points
that can be used to extract the momentum dependence of the pseudogap $\Delta$.
It is shown in Fig.~\ref{fig:fig1}(d)
that $\Delta(\pi,0)\approx30\pm5$ meV which is comparable 
with a recent ARPES report on \LBCOx\ at $x\approx1/8$ \cite{vallascience06}; 
another compound displaying a stripe phase.
Upon moving towards the zone diagonal, Fig.~\ref{fig:fig3}(a) shows
that the pseudogap remains approximately constant as a function of 
the underlying Fermi surface angle $0<\phi<15^{\circ}$ 
defined in Fig.~\ref{fig:fig2}.
Moving closer to the diagonal direction,
the pseudogap vanishes very fast and for $\phi>30^{\circ}$
gapless quasiparticles
with an enhanced lifetime are observed, 
in agreement with past studies of the pseudogap state%
~\cite{shenscience05,kanigelnature06}.

\subsection{Secondary branch}

Remarkably, a careful inspection of the spectra along cut (b) reveals
the existence of a second weaker branch of the underlying Fermi surface
in the form of a secondary peak in the MDC of Fig.~\ref{fig:fig1}(b). 
This second branch is not an artifact of a small minority 
grain since it has been observed on a handful of freshly cleaved surfaces.
It is also not a consequence of chemical disorder introduced
by Nd, since a similar secondary peak was observed in Nd-free La$_{1.88}$Sr$_{0.12}$CuO$_4$
(see later). 
Instead, we believe that this branch is intrinsic 
and that it carries important information about the electronic structure of the 
pseudogap state. 

\subsubsection{Momentum dependence of the pseudogap and the MDC-linewidth.}

\begin{figure} 
\begin{center}
\includegraphics[width=0.44\textwidth]{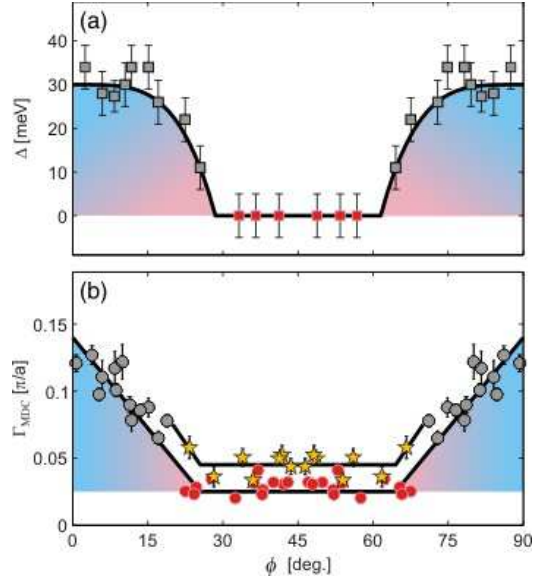}
\caption{
(a) The pseudogap $\Delta$ as a function of the polar angle 
$\phi$ defined in Fig.~\ref{fig:fig2}. 
(b) $\Gamma_{\mathrm{MDC}}$ as a function of $\phi$, 
circular and star points 
represent the primary and secondary branches, respectively. 
The solid lines are guides to the eye.
        }
\label{fig:fig3}
\end{center}
\end{figure}

Figure~\ref{fig:fig1}(e) shows a set of MDC
along cut (b) at different energies 
that reveal the opposite dispersion of the two branches.
We find from Fig.~\ref{fig:fig1}(e)
that the Fermi velocity of the primary ($v_F\approx 1.6\pm 0.05$ eV\AA)
and secondary ($v_F\approx -2.0\pm 0.3$ eV\AA) branch are consistent 
with the universal Fermi velocity along the diagonal of hole-doped cuprates \cite{zhounature03}. 
We have followed the evolution of $\bfkF$ 
and  the MDC line-width $\Gamma_{\mathrm{MDC}}$
[half width at half maximum (HWHM)]
for both  branches. The momentum dependence of the secondary branch, relative to 
that of the primary branch, is shown in Fig.~\ref{fig:fig1}(f).
The peak separation between the two branches decreases as $\phi$ decreases
from $\phi\approx 45^\circ$ to $\phi\approx 15^\circ$.
Figure~\ref{fig:fig3}(b) shows the dependence of $\Gamma_{\mathrm{MDC}}$
as a function of $\phi$.
Along the Fermi arcs 
we observe, that the line-width is approximately
constant 
$\Gamma_{\mathrm{MDC}}\approx0.03~\pi/a$ and 
$\Gamma_{\mathrm{MDC}}\approx0.05~\pi/a$ for both the primary and secondary  branches.
However, once the pseudogap opens the line-width
increases gradually to $\Gamma_{\mathrm{MDC}}\approx0.13~\pi/a$ 
for the primary branch at the zone boundary ($\phi=0^\circ$).
Here, $a\approx3.8$ \AA\ is the lattice spacing of the CuO square lattice.
The simultaneous opening of the pseudogap and the broadening of the
MDC line-width make it practically impossible to resolve the secondary branch
once it appears as a weak shoulder on the primary branch
below $\phi\approx 15^\circ$,
{\it i.e.}, close to $(\pi,0)$ [see grey circles in Fig.~\ref{fig:fig1}(f)].

The underlying Fermi surface defined by the positions of the MDC peaks
at $E_F$ is shown in Fig.~\ref{fig:fig2}. The circles identify the primary
branch, that can be followed all the way to the zone boundary.
The red circles map out four primary Fermi arcs by the criterion that gapless quasiparticle
peaks [see blue curve in Fig.~\ref{fig:fig1}(c)] are observed while
the grey points map out the segments of the underlying Fermi surface where 
the quasiparticle peaks are suppressed by the pseudogap. The gapless
quasiparticles have a finite lifetime along the primary Fermi arcs.
A more delicate analysis is needed to extract
the quasiparticle lifetimes of the secondary branch as it requires
a background subtraction in order to observe EDC peaks 
(see~\ref{app:A}). 
Here, we only display the Fermi momenta  
of the secondary branch by the stars in Fig~\ref{fig:fig2}.

\subsubsection{Temperature dependence of the secondary branch and the pseudogap.}
\begin{figure*}
\begin{center}
\includegraphics[width=0.9\textwidth]{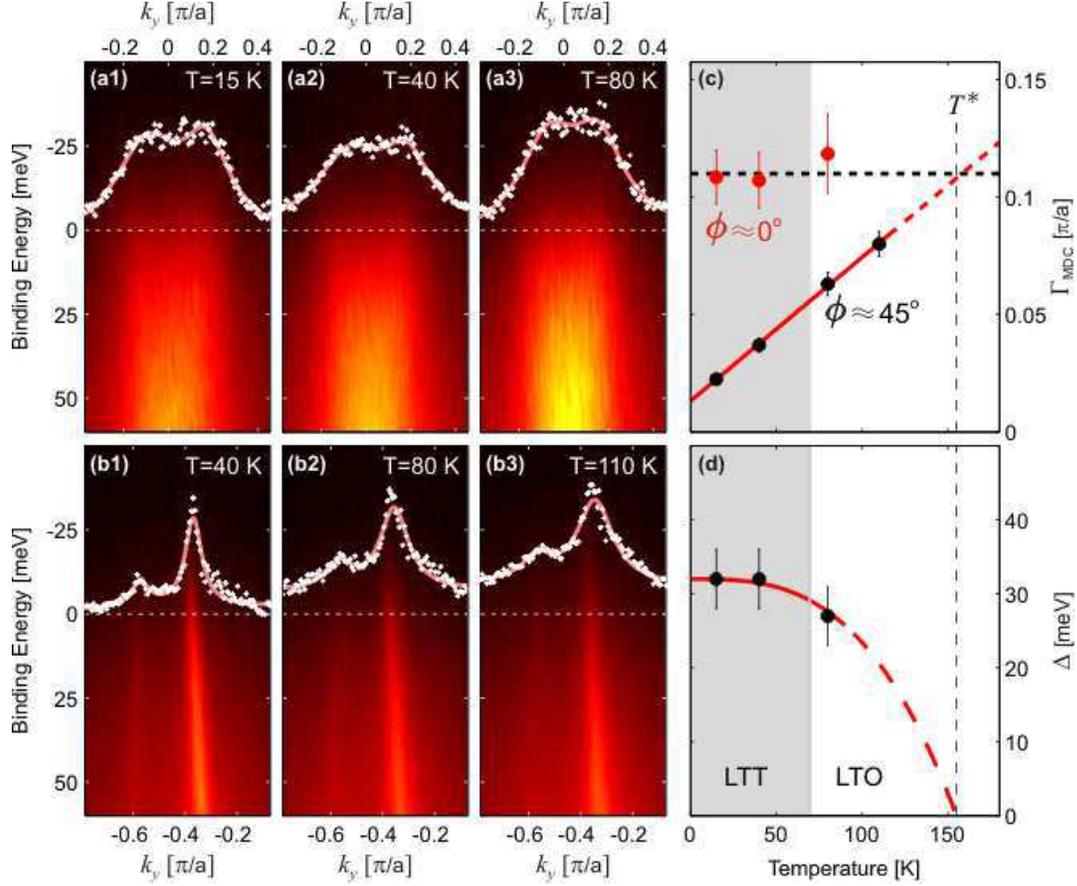}
\caption{
(a1-a3) Cut (a) of the manuscript with $T=15$, $40$, and $80$ K respectively. 
(b1-b3) are the spectrum along cut (b) of the manuscript with $T=40$, $80$, and $110$ K. 
White points are momentum distribution curves (MDC) at $\omega=E_F$ 
and the solid lines are double Lorentzian fits with a linear background. 
(c) The MDC linewidth at $\omega=E_F$ of cut (a) (red points) and cut (b) (black points)
as a function of temperature. The solid red line is a linear fit.
(d) The pseudogap $\Delta$ as a function of temperature; 
the red line is a guide to the eye. $T^*$ is defined by the derivation away from
the linear temperature dependence of the in-plane resistivity \cite{ichikawa00}.
        }
\label{fig:fig1A}
\end{center}
\end{figure*}

We now turn to the temperature dependence of the secondary branch along 
cut (b). No significant change of the intensity of the secondary 
branch for $\omega=E_F$ is observed for upon heating from $T=15$~K to $T=110$~K
(see Fig.~\ref{fig:fig1}(f)).
Figures~\ref{fig:fig1A}(a1-a3) 
show the ARPES spectra recorded along cut (a) of Fig.~\ref{fig:fig1} 
for $T=15$, $40$, and $80$ K while 
Fig.~\ref{fig:fig1A}(b1-b3)
show the spectra recorded along cut (b) 
for $T=40$, $80$, and $110$ K. 
The white points in 
Figs.~\ref{fig:fig1A}(a1-a3) 
and
Figs.~\ref{fig:fig1A}(b1-b3)
are the momentum distribution curves (MDC) at $\omega=E_F$
while the solid lines are double Lorentzian fits with a linear background.
For each MDC, the double Lorentzian fit allows to extract
the MDC linewidth $\Gamma^{\mathrm{(cut)}}_{\mathrm{MDC}}$.
The dependence on the temperature $T$ of the MDC linewidths
along cut (a), $\Gamma^{\mathrm{(a)}}_{\mathrm{MDC}}$,
and along cut (b), $\Gamma^{\mathrm{(b)}}_{\mathrm{MDC}}$,
is shown in Fig.~\ref{fig:fig1A}(c) by the red and black circles,
respectively. The temperature dependence can, 
to a first approximation, be described by
\begin{equation}\label{eq:linearfit}
\Gamma^{\mathrm{(cut)}}_{\mathrm{MDC}}(T)=
\alpha^{\mathrm{(cut)}}
+
\beta^{\mathrm{(cut)}} 
T.
\end{equation}
For cut (a) the linewidth is roughly temperature independent, thus
\begin{equation}
\alpha^{\mathrm{(a)}}\approx0.0106~\pi/a,
\qquad
\beta^{\mathrm{(a)}}\approx0,
\end{equation}
see the horizontal dashed line in Fig.~\ref{fig:fig1A}(c).
By contrast for cut (b) the linewidth exhibits a stronger 
temperature dependence and a least-square fit yields
\begin{equation}
\alpha^{\mathrm{(b)}}=0.0132~\pi/a,
\qquad
\beta^{\mathrm{(b)}}=0.0006~\pi/(a K),
\end{equation}
as shown by the solid red line in Fig.~\ref{fig:fig1A}(c).

The evolution with temperature of the spectra shown 
in Figs.~\ref{fig:fig1A}(b1-b3)
demonstrates that both the pseudogap and the secondary
Fermi arcs persist deep into the low-temperature orthorhombic (LTO) phase.
Hence, neither the pseudogap nor the secondary Fermi arcs are
directly related to the spin and charge long-range order (LRO) observed
by neutron diffraction and $\mu$SR experiments, 
as static stripes appear only
in the low-temperature tetragonal (LTT) phase%
~\cite{tranquadanature95,tranquadaprb96,christensenprl07}.

\section{Crystal structure and doping dependence}

Now we turn to discuss the possible origin of the secondary band.
Very early on the so-called shadow band was observed in Bi2212 \cite{aebiPRL94}. 
The shadow band is essentially doping and temperature independent~\cite{nakayamaPHYSC07}. 
By use of high-quality untwinned crystals and tunable light polarization~\cite{manPRL06} 
the shadow band was later shown to 
originate from the orthorhombic distorted lattice structure.
This shadow band is reminiscent of the secondary band
that we have observed in \NdLSCO (NdLSCO). 
Our NdLSCO crystals is not 
 fully untwinned. For this reason it is not possible
 to follow the experimental procedure of
 Ref.~\cite{manPRL06} to demonstrate whether the secondary branch
 in NdLSCO has it origin from a weak orthorhombic distortion.

\begin{figure*}
\begin{center}
\includegraphics[width=0.75\textwidth]{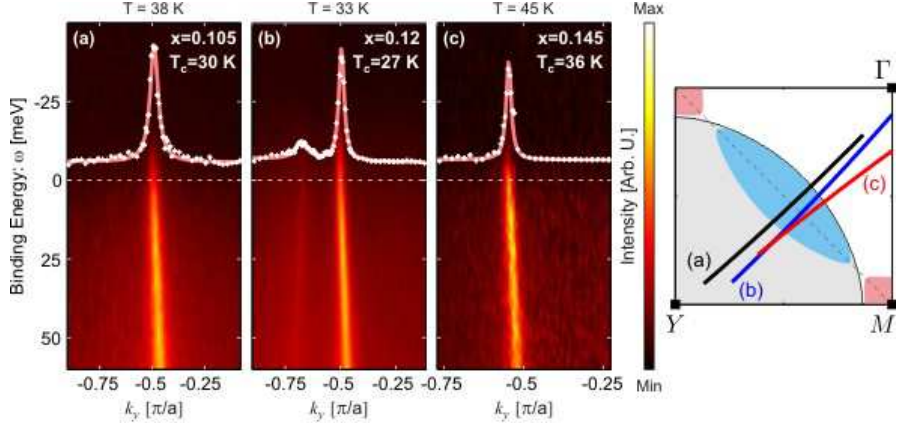}
\caption{
(a)-(c) ARPES intensity as function of binding energy and momentum 
(shown in right panel) for La$_{2-x}$Sr$_x$CuO$_4$ with $x=0.105$, $x=0.12$,
and $x=0.145$. The white points in (a)-(c) are momentum distribution 
curves at $\omega=E_F$ and the solid lines are Lorentzian fits with 
linear background. (a1)-(c1) show the underlying Fermi surface 
for  $x=0.105$, $x=0.12$,
and $x=0.145$. The blue lines indicate the momentum cuts along which the
spectra shown in (a)-(a) are recorded.}
\label{fig:dopingdep}
\end{center}
\end{figure*}

Band structure calculations generically predict that 
weak orthorhombic  distortions of a tetragonal lattice structure
lead to a band folding with $\mathbf{Q}=(\pi,\pi)$ in the tetragonal 
Brillion zone \cite{pickettRMP89}.
Accordingly one would expect to observe hole pockets 
for any weak orthorhombic distortion of a tetragonal lattice structure. 
Now, La$_{2-x}$Sr$_x$CuO$_4$  
has an orthorhombic
 lattice structure \cite{changarxiv07}
at low temperatures for $x<0.21$. Sofar there was however little
experimental evidence for a shadow band in LSCO. Nakayama {\it et al.} \cite{nakayamaPRB06}
observed a weak indication of a backfolded band in the superconducting state of 
La$_{1.85}$Sr$_{0.15}$CuO$_4$. 
X.J. Zhou {\it et al.} \cite{XJZhou} on the other hand observed the secondary band 
for $x=0.06$ and 0.09 but not for $x=0.15$. The latter study thus indicates a
rather strong doping dependence of the secondary band that does not 
correlate with an orthorhombic phase of LSCO in an obvious way. 
We now present a systematic study of the normal state electronic structure
in La$_{2-x}$Sr$_x$CuO$_4$ near the 1/8-anomaly.
In Fig.~\ref{fig:dopingdep} cuts (a)-(c), close to the zone diagonal,  
are shown for La$_{2-x}$Sr$_x$CuO$_4$  with $x=0.105$, $x=0.12$, and $x=0.145$, respectively. 
Interestingly, we observe back folding only for $x=0.12$, as shown in 
Fig.~\ref{fig:dopingdep}(b), but not for $x=0.105$ and $x=0.145$,
as shown in Figs.~\ref{fig:dopingdep}(a) and (c).
This strongly suggests that the band folding is enhanced near 
the so-called 1/8-anomaly (compounds with $x\approx1/8$). 

We now turn our attention to NdLSCO.
The average crystal structure of NdLSCO is tetragonal for $T<69$~K.
However, locally each CuO$_2$-planes
has orthorhombic distortions \cite{axe}.
Such local distortions 
could lead to a back folding \cite{norman08}. 
However, if the back folding is purely of structural origin,
one would expect a significant effect when going from the LTO to the LTT
lattice structure.
We show in Fig.~1(g) 
that the intensity of
secondary branch displays essentially no change upon going 
through the LTO-LTT transition. 

We have shown that the back folding is strongly enhanced 
near the so-called 1/8-anomaly and it appears independent 
of the (local and global) crystal structure.
We are therefore lead to the preliminary conclusion that the secondary branch 
can not be attributed solely to  orthorhombic distortions
of the tetragonal crystal structure.

\section{Discussion}
The observation of the secondary branch of excitations at the Fermi energy
is the main result of this letter. 
Whether or not the secondary branch is an effect related to the crystal structure, 
the electronic structure  can be modeled by the use of a tight-binding model.
We are now going to argue by appealing to a simple two-dimensional%
~\cite{note3}
fermiology model, 
that the primary and secondary Fermi arcs are the remnants of two hidden
Fermi pockets: a hole pocket centered at the diagonal and an electron pocket centered
at the M-point.  
The black lines in Fig.~\ref{fig:fig2} 
are the Fermi surface of the tight-binding dispersion
\begin{equation}\label{eq:eq1}
\varepsilon^{\ }_{\mathbf{k}}=
\mu
-
2t^{\ } (\cos k^{\ }_xa+\cos k^{\ }_ya)
-
4t^{\prime}\cos k^{\ }_xa\cos k^{\ }_ya,
\end{equation}
where the nearest-neighbor amplitude $t=240$ meV
while the chemical potential $\mu/t=1$
and the next-nearest-neighbor hopping amplitude 
$t^{\prime}/t=-0.325$
are fixed by the Fermi velocity along the diagonal 
and topology of the primary branch
of the underlying Fermi surface, respectively.
The Fermi surface centered around the $Y$-point has the area
\begin{equation}\label{eq:FSarea}
A_{FS}=2(1+p)A\approx2.242A
\end{equation}
where $A=(\pi/a)^2$.
Notice that the number $p\approx0.12$ of charge carriers counted from half filling  
is consistent with the nominal doping $x=0.12\pm0.005$.
Evidently, $\varepsilon^{\ }_{\mathbf{k}}$
fails to account for the secondary Fermi arc shown in Fig.~\ref{fig:fig2}.
We observe that $\bfkF$ of the primary and secondary
branch are approximately related by a reflection symmetry about
the reduced zone boundary (see dashed line in Fig.~\ref{fig:fig2}).
This is suggestive of a unit cell doubling. 
In a simple fermiology picture, such a unit cell doubling can occur
with the onset of long-range order (LRO) in a particle-hole channel
at the momentum $\mathbf{Q}=(\pi,\pi)$ that causes
the opening of a single-particle gap $\gamma$ close to the \textit{hot spots} 
 -- the crossings between the Fermi surface 
$\varepsilon^{\ }_{\mathbf{k}}=0$ and the reduced zone boundary. 
This brings about the reconstruction 
of the Fermi surface $\varepsilon^{\ }_{\mathbf{k}}=0$ 
according to~\cite{harrisonprl07} 
\begin{equation}\label{eq:eq2}
0=
\varepsilon^{\pm }_{\mathbf{k},\mathbf{Q}}\equiv
\frac{
\varepsilon^{\ }_{\mathbf{k}}+\varepsilon^{\ }_{\mathbf{k}+\mathbf{Q}}
     }
     {
2
     }
\pm
\sqrt{
\left(\frac{\varepsilon^{\ }_{\mathbf{k}}-\varepsilon^{\ }_{\mathbf{k}+\mathbf{Q}}}{2}\right)^2
+
\gamma^2
     }.
\end{equation}
By choosing the constant single-particle gap $\gamma=30$ meV
the lower branch $\varepsilon^{-}_{\mathbf{k},\mathbf{Q}}\leq0$ 
yields the hole pocket shown by the blue area in Fig.~\ref{fig:fig2}%
~\cite{note2}. There also exists a small electron pocket 
($\varepsilon^{+}_{\mathbf{k},\mathbf{Q}}\leq0$) 
centered around ($\pi,0$) as indicated by the pink shaded area 
in Fig.~\ref{fig:fig2}.
Although we do not provide any direct signature of an electron pocket,
our ARPES data are not inconsistent with an electron pocket.
It is difficult to extract such information from the 
spectrum close to the zone boundary because 
(i) the intensity is strongly suppressed at $E_F$ due to the pseudogap 
and (ii) $\Gamma_{\mathrm{MDC}}$ is comparable 
to the diameter of the predicted electron pocket.
However, transport measurements on this material%
~\cite{nakamura92,nodascience99} 
and other so-called stripe compounds%
~\cite{adachiprb01} 
have revealed the existence of a negative Hall coefficient at low temperatures;
the usual fingerprint of an electron pocket. 
Assuming that the Luttinger sum rule~\cite{luttinger} holds, it can be 
shown by simple geometrical considerations that
$p$ in Eq.~\ref{eq:FSarea} becomes
\begin{equation}
p=\left(A_{hp}-0.5A_{ep}\right)/A\approx0.12
\end{equation}
where $A_{hp}=0.155A=10.6$ nm$^{-2}$ is the area of the hole pocket and
$A_{ep}=0.064A=4.4$ nm$^{-2}$ is the volume of the electron pocket
in Fig.~\ref{fig:fig2} \cite{noteYBCO}. 
The naive two-band structure, suggested in Eq.~\ref{eq:eq2}, is one
way to reconcile  Hall-coefficient and ARPES experiments. Sofar there
exist, however, little spectroscopy evidence for a two-band structure
in hole doped cuprates. For electron doped cuprates there exist on 
the other hand several reports claiming the observation of a two-band
structure for dopings slightly larger than $x\sim1/8$ \cite{matsuiPRL05,parkPRB07}.

The qualitative observation of a hole pocket is compatible with many theories. 
Band structure calculations have shown 
that pockets can form in orthorhombic 
crystal structures \cite{pickettRMP89}. 
However, the fact that the secondary band is enhanced
for $x\approx0.12$ in orthorhombic La$_{2-x}$Sr$_{x}$CuO$_4$ 
would suggests that orthorhombic lattice distortions can  not be 
the primary cause for the Fermi surface reconstruction that we observe. 
Models that invoke spin and charge separation
of the electron quantum numbers%
~\cite{lee08,sachdevnatphys08,ricePRB06,senthil08} 
or competing orders%
\cite{chubukov97,harrisonprl07,millisnorman07,chakravartyarxiv07,granathPRB08} 
also predict a hole pocket in the pseudogap phase. 
Moreover, 
an electron pocket is predicted to coexist with a hole pocket
as a result of competing orders%
~\cite{chubukov97,harrisonprl07,millisnorman07,chakravartyarxiv07}. 
A clue favoring the scenario by which competing order is
causing the Fermi-surface reconstruction seen by ARPES in
NdLSCO and La$_{1.88}$Sr$_{0.12}$CuO$_4$ is the observation of a 
spin-density wave LRO in these materials \cite{changarxiv07}. 
However,
the fact that the Fermi surface reconstruction 
is present above the onset temperature of the spin-density wave LRO suggests that this 
true LRO is not required for the formation of pockets~\cite{chubukov97,harrisonprl07}.

\section{Conclusions}
In summary, our unambiguous observation of a secondary branch
of gapless quasiparticles around the zone diagonal 
in the stripe-compound NdLSCO
is consistent with a hole-like Fermi surface
induced by a unit cell doubling with the characteristic momentum
$\mathbf{Q}=(\pi,\pi)$.
Such a secondary branch was also observed in underdoped LSCO ($0.03<x<0.06$)
by X.J. Zhou {\it et al.} \cite{XJZhou}. Our results on Nd-free LSCO 
suggest that this secondary branch is enhanced near the so-called
1/8-anomaly. While the results around the 1/8-anomaly might suggest a connection 
to a spin density wave effect, the non-monotonic doping dependence implies 
a non-trivial interplay of spin and lattice degrees of freedom. 

\ack
This work was supported by the Swiss NSF 
(through NCCR, MaNEP, and grant Nr 200020-105151, PBEZP2-122855), the Ministry
of Education and Science of Japan and the Swedish Research Council.
This work was entirely performed at the Swiss Light Source
of the Paul Scherrer Institute, Villigen PSI, Switzerland. 
We thank the beamline staff of X09LA for their support and
we thank M.~R. Norman, T.~M. Rice, and F.~C. Zhang for discussions.

\appendix
\section{Quasiparticles on the secondary branch}\label{app:A}
\begin{figure}
\begin{center}
\includegraphics[width=0.49\textwidth]{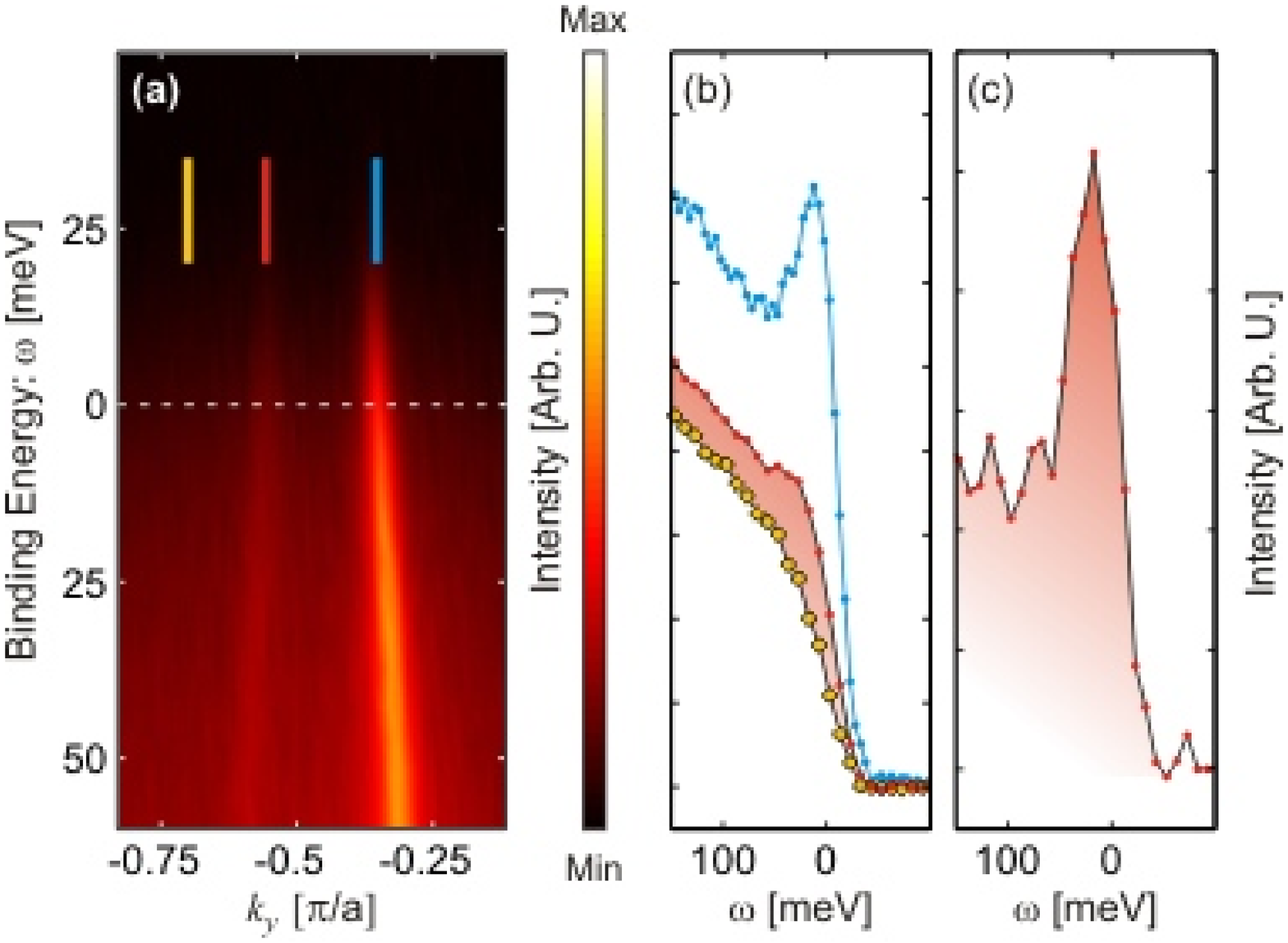}
\caption{
(a) The same ARPES spectrum as shown in Fig.~1(b) of the manuscript. 
(b) The blue and red points are the energy distribution curves at $\mathbf{k}_F$ 
located at the primary and secondary Fermi arcs, respectively. 
The yellow points are a typical energy distribution curve that 
never crosses the primary
and secondary dispersing branches seen in (a).
(c) Difference between the red and yellow points in (b).
        }
\label{fig:figEDC}
\end{center}
\end{figure}

It was shown in the manuscript that gapless spectral peaks 
are visible along the primary Fermi arc centered
around the zone diagonal. We are now going to examine more
closely the spectral peaks in the energy distribution curves (EDC)
associated to the secondary Fermi arc.  
Figure~\ref{fig:figEDC}(a) shows the spectrum 
along cut (b) at $T=15$ K. 
Figure~\ref{fig:figEDC}(b) shows the EDC at $\mathbf{k}_F$ 
[determined by the 
MDC at $\omega=E_F$ and indicated by the blue and red lines in \ref{fig:figEDC}(a)]
for both the primary and secondary Fermi arcs with blue and red points, respectively. 
The yellow points sample an EDC with $\bfk$ chosen sufficiently
far from the detector edge but still not too close to the Fermi arcs
so as not to cross either the primary or the secondary dispersing branches seen in (a).
This EDC may therefore represent an average background intensity observed by ARPES.
To better visualize the EDC at $\mathbf{k}_F$ for the secondary Fermi arc, we subtract 
in Fig.~\ref{fig:figEDC}(c) the yellow data points from the red data points
in Fig.~\ref{fig:figEDC}(b). Although a peak remains, it is difficult to analyze it
so as to extract a quasiparticle lifetime. We thus limit our conclusions to 
the fact that we observe a weak but significant spectral peak at $\bfk_{F}$ 
on the secondary branch.   

Finally we comment on the fate of the secondary branch in the 
superconducting state of NdLSCO. Since the onset temperature of superconductivity
in our NdLSCO sample ($T_c\sim7$ K) is below the lowest accesible temperature 
of the SIS instrument, we have not been able to study the 
secondary branch of NdLSCO in the superconducting state.
The effect on the secondary branch upon cooling into the superconducting 
state of NdLSCO remains therefore an outstanding open problem.
However, we were able to confirm that the secondary branch persist into
the superconducting state of La$_{1.88}$Sr$_{0.12}$CuO$_4$  although the applied energy resolution
did not permit the determination of a superconducting gap on the 
secondary branch. 

\section{Temperature dependence of the secondary branch}\label{app:B}
\begin{figure}
\begin{center}
\includegraphics[width=0.45\textwidth]{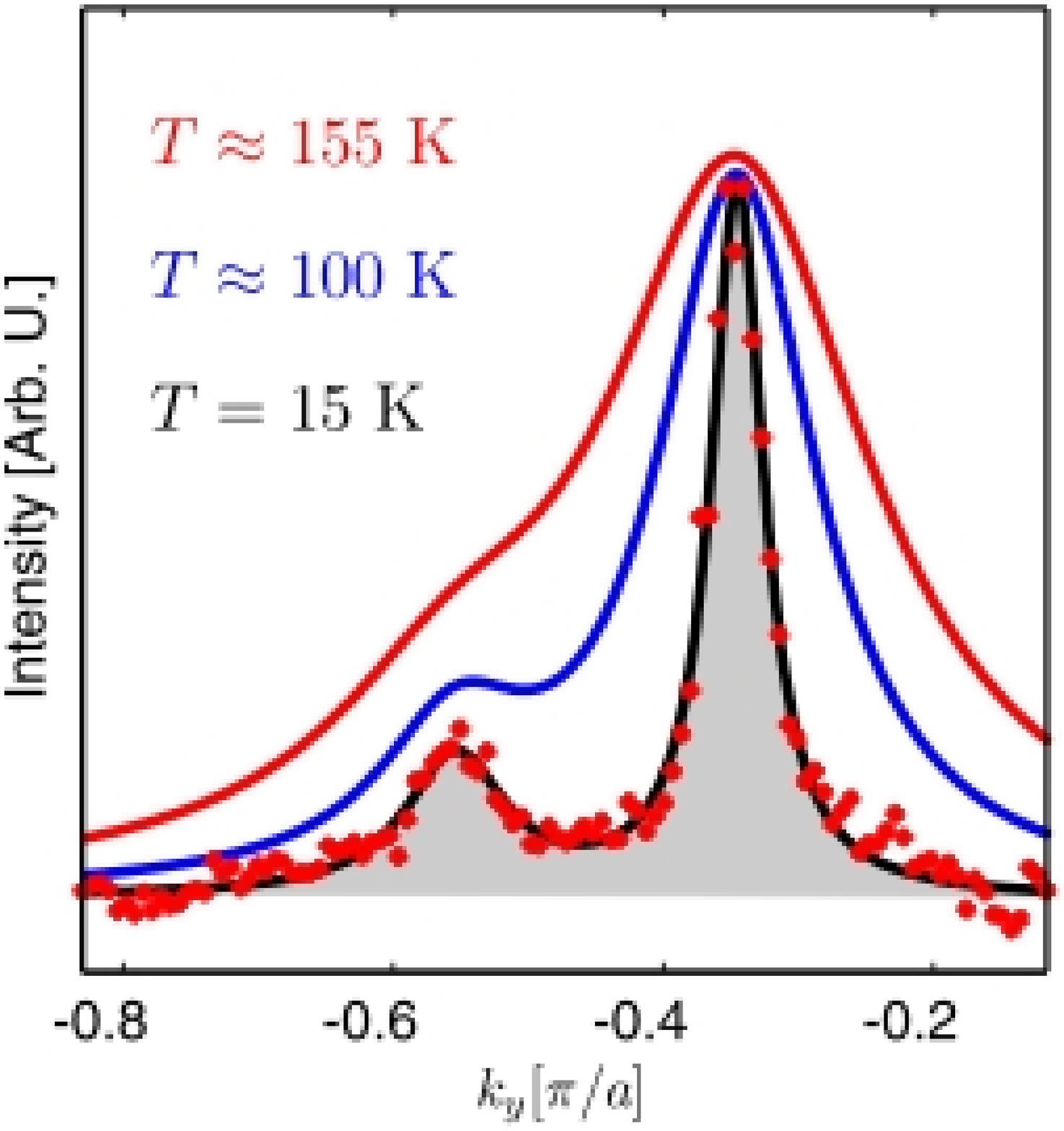}
\caption{
The measured red points sample the MDC at $\omega=E_F$ and $T=15$ K for the cut (b) 
while the black line is a double Lorentzian fit with a linear background.
The blue and red curves are the extrapolated MDC at $T=100$ K and $T=155$ K,
respectively, assuming a double Lorentzian fit with a linear background
whereby the maximum is taken to be temperature independent while the linewidth
increases according to Eq.~(\ref{eq:linearfit}).
        }
\label{fig:figT}
\end{center}
\end{figure}

What could be the energy (temperature) scale that controls the existence
of the secondary Fermi arcs? We have argued in the manuscript on the basis
of a simple fermiology model that the pseudogap energy scale might be related
to the existence of a hole-like Fermi pocket. If so, we would expect that
the secondary Fermi arc should disappear with increasing temperature at
the temperature $T^{*}\approx 155$ K,
below which the pseudogap manifests itself
(as measured from the deviation away from
the linear temperature dependence of the in-plane resistivity \cite{ichikawa00}).
This expectation is consistent with the observation that the
linear fits along cuts (a) and (b) in Eq.~(\ref{eq:linearfit})
intersect at the temperature $T^{*}\approx 155$ K
as shown in Fig.~\ref{fig:fig1A}(c).
Unfortunately, the direct experimental verification of this
educated guess is ambiguous due to the thermal broadening of the MDC along cut
(b). This is illustrated in
Fig.~\ref{fig:figT} where we plot the double Lorentzian fits 
to the MDC along (b) at the Fermi energy with their
temperature-dependent widths taken from Fig.~\ref{fig:fig1A}(c),
assuming a temperature independent peak amplitude. Whereas 
it is still possible to extract $\mathbf{k}_F$ and
the linewidth of the secondary branch
at $T\approx110$ K this task becomes ambiguous at
 $T\approx T^*\approx155$ K as the secondary Fermi arc is signaled
by a small shoulder on the dominant peak induced by the primary Fermi arc
(see Fig.~\ref{fig:figT}).

\newpage


\begin{thebibliography}{99}
\bibitem{note3}
Since the out-of-plane resistivity $\rho_c$ is three orders of magnitude 
larger than the in-plane resistivity $\rho_{ab}$ 
in this compound \cite{tranquadaprb96} 
the Fermi surface must be predominantly two-dimensional.
The herein applied fermiology model neglects electronic interactions
beyond the existence of an energy gap 
and assumes translation invariance (no impurities).
\bibitem{noteYBCO}
Notice that $A_{ep}$ compares well with the
 volume of the electron pocket $A_{ep}=5.1$ nm$^{-2}$ observed by quantum oscillations in
 YBCO \cite{leyraudnature07}.
\end{thebibliography}
\end{document}